\documentclass[aps,prl,twocolumn,groupedaddress]{revtex4}
\usepackage{graphicx}
\usepackage{amsmath}
\usepackage{color}

\begin{document}

\title{A metamaterial analog of electromagnetically induced transparency}

\author{N. Papasimakis}
\affiliation{Optoelectronics Research Centre, University of
Southampton, SO17 1BJ, UK} \email{N.Papasimakis@soton.ac.uk}

\author{V. A. Fedotov}
\affiliation{Optoelectronics Research Centre, University of
Southampton, SO17 1BJ, UK}

\author{S. L. Prosvirnin}
\affiliation{Institute of Radio Astronomy, National Academy of
Sciences of Ukraine, Kharkov, 61002, Ukraine}

\author{N. I. Zheludev}
\affiliation{Optoelectronics Research Centre, University of
Southampton, SO17 1BJ, UK}

\date{\today}

\begin{abstract}

We present a new type of electromagnetic planar metamaterial that
exhibit strong dispersion at a local minimum of losses and is
believed to be the first metamaterial analog of electromagnetically
induced transparency. We demonstrate that pulses propagating through
such metamaterials experience considerable delay, whereas the
thickness of the structure along the direction of wave propagation
is much smaller than the wavelength, which allows successive
stacking of multiple metamaterial slabs. This leads to a significant
increase in the band of normal dispersion, as well as in
transmission levels.
\end{abstract}

\maketitle

Changes in the velocity of light propagating through dispersive
media have been the subject of extensive investigations in the past
\cite{Brill}. In fact, it was predicted \cite{Garret} and later
observed \cite{Casper,Chu} that light pulses can propagate with
apparent velocities greater or smaller than in vacuum without strong
distortion of shape and width due to pulse reshaping phenomena.
Today, control over these effects is essential for the development
of optical communication technologies leading to a number of
different approaches \cite{EIThau99, CRYSTturukh01, CRYSTboyd03,
PCscalora96, PCnotomi01, PCvlasov05, QWku04, FIBgaeta05, FIBsong05,
SOLmok06, SPPkuip07, MMWGhess07}, such as the quantum phenomenon of
electromagnetically induced transparency (EIT)
\cite{EITharris90,EITstoppedhau01}. In that case, an otherwise
opaque atomic medium is rendered transparent in a narrow spectral
region within the absorption line through quantum interference of a
pump and a probe laser beam \cite{EITharris90}. This sharp
dispersion has important consequences such as dramatically reducing
the group velocity \cite{EIThau99} and enhancing non-linear
interactions \cite{EITnonlin}. Most of the proposed techniques,
however, involve special experimental conditions, e.g. cryogenic
temperatures, coherent pumping and high intensities.

Recently, the implementation of EIT-like behavior in classical
systems has attracted a lot of attention \cite{cEIT0, cEIT1, cEIT2,
cEIT3, cEIT4, cEIT5, cEIT6, cEIT7, cEIT8}, since in this case the
operating frequency can be tuned simply by varying the system
geometry, while no pumping is necessary. In particular, interference
between coupled classical resonators can lead to the same effects as
EIT, i.e. narrow transmission resonances within the single-resonator
stop-band. Nevertheless, existing approaches involve spatial
dispersion at the wavelength scale, introducing, therefore,
fundamental constraints on the minimum thickness of the medium. In
this letter, we demonstrate a classical analogue of EIT in planar
metamaterials, namely metal-dielectric slabs of vanishing thickness
along the propagation direction, periodically patterned on a
subwavelength scale. The phenomenon arises as a result of engaging
"trapped-mode" resonances that are weekly coupled to the free space
leading to low transmission losses and exceptionally high quality
factors. Such modes are normally forbidden, but can be excited in
planar metamaterials with special patterning \cite{darkth}.

\begin{figure}[h] \label{fig1}
\includegraphics[width=.45\textwidth]
{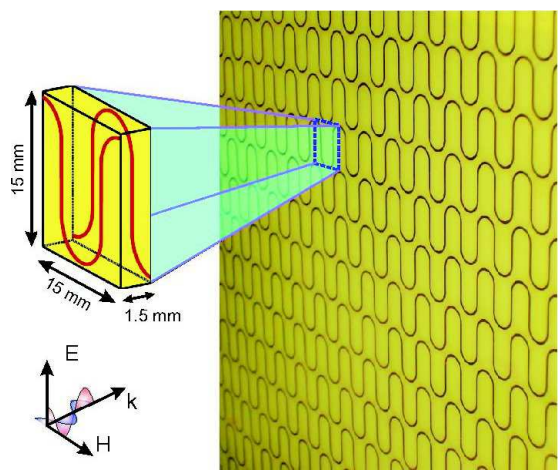}\caption{The bilayered fish-scale metamaterial and its
unit cell. The two fish-scale patterns (red) reside in the top and
bottom face of the dielectric, respectively, and are displaced along
the meandering stripes by a half unit cell in respect to one
another.}
\end{figure}

The studied metamaterial is based on the continuous fish-scale
copper pattern, which is known for its resonant properties
\cite{fs}. To illustrate the difference between a conventional
planar frequency selective structure and the suggested EIT-like
metamaterial, we manufactured two types of structures. In the
reference structure, the metallic pattern is etched on one side of
the dielectric slab, while in the EIT-like metamaterial the pattern
resides on both sides of the PCB laminate, so that the pattern on
one side of the dielectric slab is shifted along the meandering
strips by half a translational cell with respect to the pattern on
the other side (see Fig. 1). As it will be shown, this inversion of
the second layer combined with the separation of the two layers
leads to strong confinement of electromagnetic energy in the gap
between the two layers and, consequently, to significantly different
properties than those of the reference structure. Both metamaterials
were manufactured by etching $35~\mu m$ thick copper films on
fiberglass PCB substrates of $1.5~mm$ thickness. The size of the
translational cell was $15~mm \times 15~mm$ rendering the
metamaterials non-diffracting at frequencies below $20~GHz$.

\begin{figure}[!h]
\label{fig2}
\includegraphics[width=.5\textwidth]{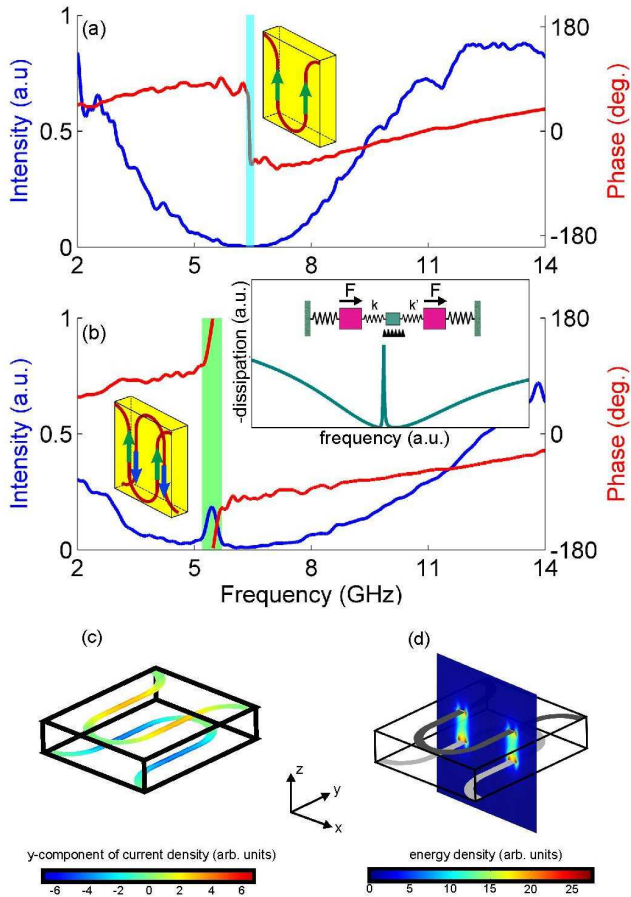} \caption{Transmission amplitude (blue) and phase
change (red) of the single (a) and the bilayered (b) fish-scale
metamaterial. Frequency regions of resonant anomalous (a) and normal
(b) dispersion are highlighted. In both cases the polarization is
along the vertical axis as shown in the inset to Fig. 1. The current
configurations are marked by the blue and green arrows in the insets
showing the samples. In the inset to Fig. 2b, the mechanical analog
of the bilayered fish scale and its response are shown. Numerically
simulated current and energy density at the resonant frequency
($5.5~GHz$) are shown in (c) and (d), respectively.}
\end{figure}

The transmission properties of the metamaterials were measured in an
anechoic chamber at normal incidence, in the frequency range of $2$
to $14~GHz$, using two broadband horn antennas and a vector network
analyzer. The measured spectra for the single- and double-layer fish
scales are presented in Figs. 2a and 2b, respectively. In the
reference case of the single fish-scale structure, a wide stop band
is visible, centered at $6.5~GHz$ and accompanied, as the
Kramers-Kronig relations dictate, by \textit{anomalous} dispersion.
On the contrary, when the bilayered metamaterial is considered (Fig.
2b), the spectrum exhibits a narrow transmission resonance centered
at around $5.5~GHz$, in the middle of the stop band of the reference
structure. The bandwidth of the transparency window is $0.5~GHz$,
while the transmission level exceeds $20\%$. Moreover, the quality
factor of this resonance is significantly higher ($>10$) than what
is typically encountered in frequency selective surfaces. As a
result, very sharp \textit{normal} dispersion is also observed,
which can lead, subsequently, to long pulse delays.

The behavior of the double layer fish-scale is a result of "trapped
mode" excitation that is weakly coupled to free space radiation and
originates from the special patterning of the metamaterial. More
concisely, as our numerical simulations show, the currents in the
two fish-scale layers oscillate with opposite phases (Fig. 2c),
leading to concentration of electromagnetic energy in the region
between the overlapping segments (Fig. 2d). In the far-field zone,
the waves emitted by these resonant, antisymmetric currents
interfere destructively, hence ensuring the high q-factor of the
resonance by dramatically reducing scattering, the principal
mechanism of losses in metamaterials at this frequency range.
Furthermore, since the studied metamaterials exhibit no diffraction
losses in the measured spectral range, the observed losses at the
transmission peak can be attributed mainly to energy dissipation in
the dielectric (that may be reduced with a higher quality
substrate), and secondarily to the weak scattering that occurs at
the non-overlapping curved segments of each fish-scale layer.

The response of the bilayered fish-scale metamaterial is a direct
classical analog of EIT, as the weak coupling of the
counter-propagating currents to free space in the metamaterial is
reminiscent of the weak probability for photon absorption in EIT
\cite{EITharris90}. The main difference lies in the fact that, in
the present case, the transparent state is a result of classical
field interference, rather than quantum interference of atomic
excitation pathways. Moreover, in the EIT-like metamaterial system
no external pumping is required. In fact, similarly to how an EIT
system may be modeled by a set of interacting classical harmonic
oscillators \cite{cEIT0}, the response of the bilayered metamaterial
may also be explained by appealing to coupled classical oscillators:
the currents on the two fish-scale layers can be represented by two
oscillating masses on elastic springs coupled through much softer
springs to a third lighter mass, which accounts for the far-field
interference of the two layers (see inset to Fig. 2b). In this
analogy, the small mass is subject to friction, which represents the
scattering losses of the metamaterial, while the two large mass
oscillators are lossless. The system is excited by an external force
acting on both large mass oscillators, representing the incident
electromagnetic wave. When the oscillators are identical, the
mechanical system supports only symmetric modes leading to
high-amplitude oscillations of the small mass and consequently to
high dissipation. However, the situation can change by introducing a
small asymmetry, for example by allowing different stiffness in the
springs coupling the large masses to the smaller one (see inset to
Fig. 2). This allows for the antisymmetric mode to be excited, where
the two large masses oscillate with large amplitudes and opposite
phases, while the small mass remains still. All the energy pumped by
the external force is being stored in the oscillations of the large
masses and the dissipative losses in the system are thus minimized.
In the metamaterial case, the excitation of the "trapped mode" is
achieved by the relative displacement of the two layers along the
propagation direction, which leads to a phase difference in their
excitation by the incident electromagnetic wave. In fact, the
response of both the metamaterial and the oscillator system are very
similar exhibiting high quality factor resonances, as shown in the
inset to Fig. 2b. However, in the first case, the resonance is
broadened by additional (dissipative and radiative) losses resulting
in a weaker transmission peak.

\begin{figure}[!h]
\label{fig3}
\includegraphics[width=.5\textwidth]
{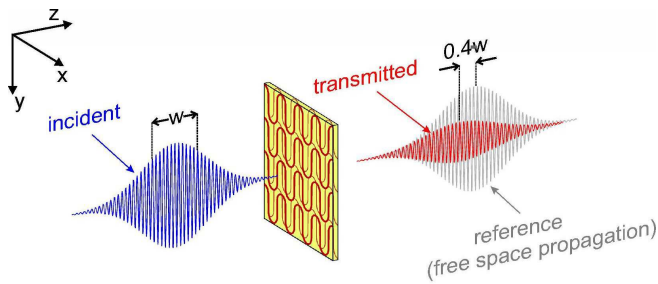} \caption{Reconstructed response of the bilayered fish
scale (red) to a $2.5~ns$ microwave Gaussian pulse (blue) with
center frequency at the maximum of the transmission window. The
transmitted pulse is delayed by $40\%$ of the pulse width, when
compared to a pulse propagating through free space (grey).}
\end{figure}

The EIT-like nature of the meta-material response  can be further
illustrated by considering propagating electromagnetic pulses,
rather than monochromatic plane waves. Indeed the ability to delay
pulses substantially with only small attenuation and minimal
distorsion is a most characteristic and useful property of EIT
media. To this end, we reconstruct the response of the metamaterial
by applying the inverse Fourier transform to the convolution of the
transmission spectrum with the pulse power spectrum. In particular,
we consider Gaussian-shaped pulses with $2.5~ns$ width at half
maximum. The center frequency of the pulse is near the peak of the
transmission band, while its spectral width is roughly equal to that
of the pass band. As a result of the strong normal dispersion, a
microwave pulse propagating through the bilayered fish-scale array
will be delayed by $\sim1~ns$, which is greater than $1/3$ of the
incident pulse width (see Fig. 3). At the same time, the pulse
retains its Gaussian shape with the exception of a weak broadening,
since most of its spectral power lies in regions where the phase
derivative is approximately constant. Achieving this level of pulse
delay is remarkable when one considers the thickness ($\lambda/35$)
of the structure. Moreover, the transmission is reasonably high,
exceeding $\sim15\%$ (may be improved by using substrates with lower
losses), and enables the successive stacking of metamaterial slabs,
at the expense, however, of pulse intensity.

Indeed, we demonstrated that the fish-scale design can be cascaded
by stacking multiple layers in such a way, that each slab is
inverted with respect to the adjacent ones (see insets to Fig. 4),
while the distance between successive fish-scale patterns is defined
by the thickness of the dielectric substrates ($1.5~mm$). The
experimental results for three and four fish-scale layers are shown
in Figs. 4a and 4b, respectively. In both cases, stacking of
multiple slabs results in an \textit{increase} of maximum
transmitted intensity and width of the frequency region, where
normal dispersion occurs. More concisely, with each additional
layer, a new transmission peak appears shifted at lower frequencies
with respect to the bilayered case. At the same time, the sharp
normal phase dispersion extends over a wide frequency band exceeding
$1.5~GHz$ and $2.5~GHz$ for three and four layers, respectively.
This allows to address the frequent requirement of practical
applications for high bandwidth, whereas the thickness of the
resulting metamaterial remains much smaller than the wavelength (at
least $\lambda/10$ in the present case).

\begin{figure}[!h]
\label{fig4}
\includegraphics[width=.5\textwidth]{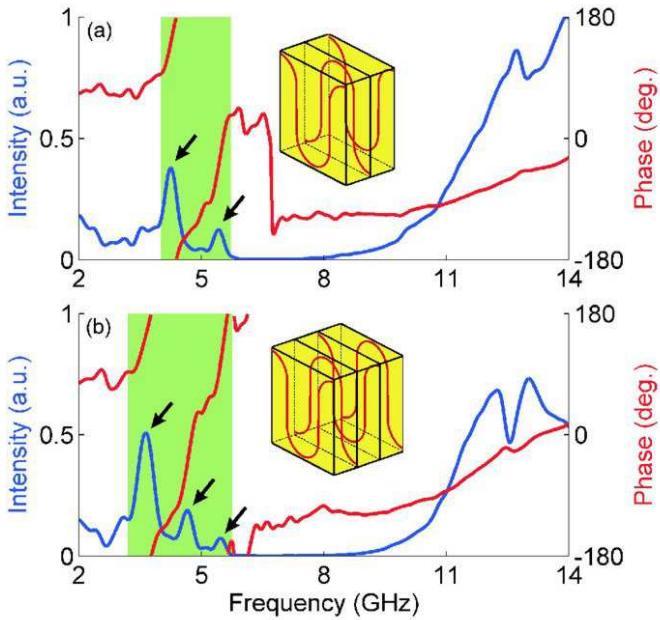} \caption{Transmission amplitude (blue) and phase
change (red) for three (a) and four (b) fish-scale layers. The
arrows mark the positions of individual resonances, while the area
of normal dispersion is highlighted (green). The polarization is
along the horizontal axis as shown in the inset to Fig. 1.}
\end{figure}

In conclusion, we propose a new type of type of planar
meta-material, which shows behavior analogous to that of EIT media,
enabling thus long optical delays in very thin, sub-wavelength
structures. This approach allows successive stacking of metamaterial
layers, in order to increase transmission and bandwidth. These
properties make such metamaterials particularly appealing as
broadband, ultra-compact delay lines operating at a prescribed
wavelengths.

\begin{acknowledgments}
The authors would like to acknowledge the financial support of the
Engineering and Physical Sciences Research Council, UK.
\end{acknowledgments}

\end{document}